\documentstyle[12pt,twoside,fleqn,espcrc1,epsfig,rotate]{article}

\newcommand{\AmS}{{\protect\the\textfont2
  A\kern-.1667em\lower.5ex\hbox{M}\kern-.125emS}}

\title{Microscopic dynamics of a phase transition: equilibrium vs out-of-equilibrium regime}

\author{V.Latora and A. Rapisarda
\address{Dipartimento di Fisica e Astronomia, 
                    Universit\'a di Catania,\\
        and INFN sezione di Catania, Corso Italia 57, I-95129 Catania, Italy}%
       \thanks{Talk given by A. Rapisarda. E-mail addresses: 
           vito.latora@ct.infn.it, andrea.rapisarda@ct.infn.it}
}
       
\begin{document}

\maketitle

\begin{abstract}
We present for the first time to the nuclear physics community the Hamiltonian 
Mean Field (HMF) model. The model can be solved  analytically in the canonical ensemble 
and shows a second-order phase transition in the thermodynamic limit. 
Numerical microcanonical simulations show interesting features in the out-of-equilibrium 
regime: in particular the model has a  negative specific heat. 
The potential relevance for nuclear multifragmentation is discussed.
\end{abstract}

{\it Invited talk at the Int. Conference CRIS2000,''Phase transitions in strong interactions: status and perspectives'', Acicastello, Italy, May 22-26 2000. }

\section{INTRODUCTION}

In the last years there has been a lot of interest in the nuclear 
physics community for multifragmentation reactions and for the study of 
liquid-gas phase transition in finite systems such as nuclei 
\cite{aladin,indra,chi,eos,eos2,pagano}. 
It is still a debated question whether nuclear multifragmentation is an 
equilibrium type of transition, and if it is first or second-order 
\cite{gross,bond,aldo,cmd,ata,campi,dago,bauer,moretto}. 
With the scope of adding new useful arguments to this debate, 
we present for the first time to the nuclear physics community 
the Hamiltonian Mean Field (HMF) model, a 
system of classical spins coupled through 
long-range attractive forces. 
HMF can be solved analitically in the canonical ensemble 
and shows a second-order phase transition in the thermodynamic limit.
Moreover HMF has the double advantage of allowing  a {\em microcanonical} 
and {\em  dynamical} approach to explore the dynamics of a phase transition  
in a finite system.
We believe that, in the same spirit of the Ising model, introduced 
many years ago, but still studied with extreme interest 
in statistical mechanics, the HMF model, though its dynamics is probably  
less complex than that one of a real system, 
can become a tool of primary importance to extract 
information on the microscopic dynamics of a second-order phase transition. 

In this paper we discuss the equilibrium and out-of equilibrium properties of 
the model, focusing our attention 
on the importance of the relaxation to equilibrium and 
on the possible discrepancies between different ensembles. 
In this respect the model can be extremely
useful also for the nuclear multifragmentation phase transition. In 
particular in section 3, we show how misleading information on the order 
of a phase transition can be obtained from a specific heat analysis \cite{dago} 
if the system is not perfectly equilibrated.

\section{HMF: THE CANONICAL ENSEMBLE}

The Hamiltonian Mean Field (HMF) model was introduced 
by Ruffo \cite{ruffo} in 1994 and has been since then 
intensively studied both analitically and numerically
\cite{latora98,latora99,prl99,pro2000,ruffod}. 
The Hamiltonian of the model describes a system of N fully-coupled particles  
moving on the unitary circle, where the particles are characterized by the angles $\theta_i$ 
and the conjugate momenta $p_i$. It reads:
\begin{equation}
    H(\theta,p)  ~  = ~ K + V ~~=
                       ~\sum_{i=1}^N  {{p_i}^2 \over 2} ~~+~~ 
 {1\over{2N}} \sum_{i,j=1}^N  \left[ 1-cos \left( \theta_i -\theta_j \right) \right]~~,
\end{equation}
\noindent
 being $K$ and $V$ the kinetic and potential energy.
If one considers a spin vector associated to each particle
${\bf m}_i \equiv [cos{\theta_i}, sin{\theta_i}]$, 
the Hamiltonian then describes a linear chain of 
$N$ classical fully-coupled spins, similarly to the XY model, and 
we can define a total magnetization vector 
${\bf M} \equiv [M_x,M_y] = {\frac{1}{N}}\sum_{i=1}^N {\bf m}_i$ .
%
%
\begin{figure}
\begin{center}
\epsfig{figure=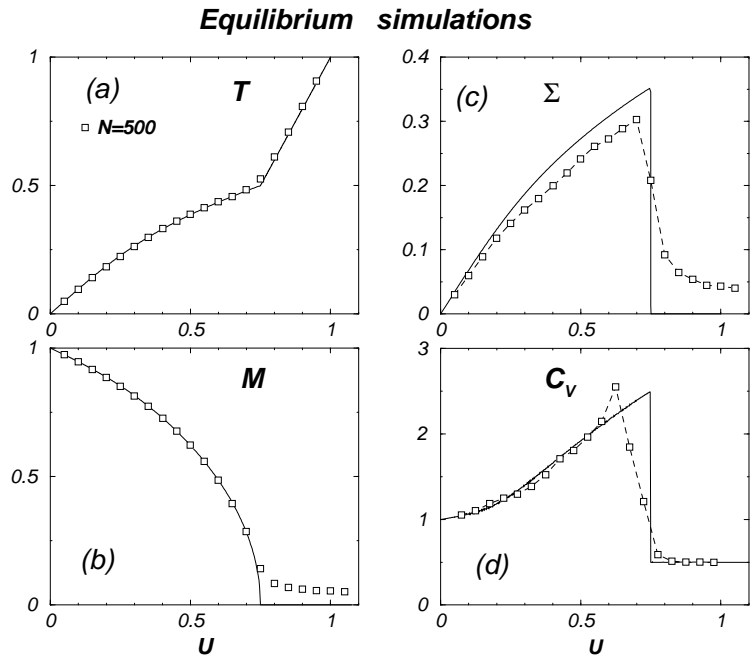,width=9truecm,angle=-90}
\end{center}
\caption{ Equilibrium microcanonical simulations for N=500 (open squares) 
in comparison with the theoretical predictions (full curves). The numerical
results are averaged over several runs. 
The small discrepancies around the critical point ($U_c=0.75$) 
and in the high energy region are due to finite-size effects.}
\end{figure}
%
A very important feature of the HMF model is that it presents an  
{\it analytical solution in the canonical ensemble}. 
The solution has been worked out by Antoni and Ruffo, 
by means of the Hubbard-Stratonovich method to perform the integration 
over the angles\cite{ruffo}. 
The system behaves as a ferromagnet at low energy and 
shows a second-order phase transition at the critical 
energy density $U_c=0.75$  ($U=E/N$ where $E$ is the total energy), 
corresponding to a critical temperature $T_c=0.5$~\cite{ruffo,latora98,latora99}. 
If the sign of the interaction is inverted, one gets an antiferromagnetic behavior: 
this case has been studied in detail in refs ~\cite{ruffo,pro2000,ruffo2000}. 
The theoretical caloric curve reads: 
\begin{equation} 
U={E\over N}={\partial (\beta F) \over \partial \beta }  =  {1 \over (2\beta) }
          + {1\over 2} \left(1 -M^2 \right)  
~~,
\label{can} 
\end{equation} 
where $\beta=1/kT$ is the inverse of the temperature 
(the Boltzmann constant $k$ is set to be equal to $1$).
The canonical caloric curve is shown in fig.1 (a) as a full curve: it increases almost
continuously with energy, but   
has a discontinuity at the critical point. The modulus of the total magnetization, $M$,
shown as full curve in fig.1(b), is close to one for very 
low energy and decreases 
to zero by increasing the energy. It is equal to zero for $U \ge U_c$. 
It has been shown that the phase-transition in the HMF model is a second-order one 
with mean field critical exponents $\beta=1/2$, $\alpha=0$ ~\cite{ruffo,latora99}. 
It is very intriguing  the fact that, though HMF was 
not constructed 
to describe a nucleus, its caloric curve  turns out to be 
very similar to the one measured by the ALADIN group\cite{aladin} 
 and more recently by other experimental 
collaborations\cite{indra,chi,eos2} for nuclear systems.

\section{HMF: THE MICROCANONICAL ENSEMBLE}

In this section we discuss the microcanonical numerical simulations and we compare 
with the equilibrium canonical solution. 
The Hamilton equations of motion for the $N$ particles are  
\begin{equation} 
\dot{\theta_i}={p_i}~~, ~ ~ ~ ~ ~~~\dot{p_i}  = -sin(\theta_i) M_x  + cos(\theta_i) M_y  
~~,~~~ 
i=1,...,N~~,
\label{eqmoto2} 
\end{equation}
and can be solved numerically for different sizes of the system. 
In this way HMF allows a {\it dynamical microcanonical approach}, perfect to 
explore the dynamics of a phase transition in a finite system. 
The equations of motion were integrated by means of a 4th order 
simplectic  algorithm~\cite{yoshida} with a time step $\delta t=0.2$
and a relative error in the total conserved energy  smaller than 
$\Delta E/E=10^{-5}$. More  details can be found in Refs~\cite{latora98,latora99}.

\subsection{Equilibrium Results}

In this paper we consider a system with $N=500$, 
a number of particles of the same order of the typical number of nucleons 
in a nuclear multifragmentation reaction. 
We start the system out-of-equilibrium, in the so-called ``water bag'', i.e. 
by putting all rotators at $q_i=0$ and giving an initial velocity 
according to a constant probability distribution function of finite 
width centered around zero \cite{ruffo,latora98}. 
We remind the reader that, though the interaction in HMF is  not explicitly 
constructed to model a nuclear system, such a far-off-equilibrium initial 
condition is a good way to simulate the strong off-equilibrium conditions 
present in a hot and compressed nuclear system before multifragmentation. 

We follow the dynamics of HMF up to a final  time  $t=2\cdot 10^6$
and we show  in fig.1 the final results of the numerical integration. The numerical 
results were integrated also on several different runs.
Such long integration times are necessary because, when out-of-equilibrium initial 
conditions are used, the system equilibration can be very slow. This is especially 
evident close to the critical point, where the presence 
of quasi-stationary (long living) non-equilibrium states
(QSS)\cite{latora98,latora99} has been revealed. 
The temperature, the total magnetization, the 
kinetic fluctuations and the specific heat per particle are reported  
as a function of the energy density as open squares and compared to 
the theoretical predictions (full curves).   
The temperature is computed from the average kinetic energy per particle  
$T=2<K>/N$, where the symbol $<~>$ stands for time averages. 
The kinetic energy fluctuations $\Sigma$, obtained from 
the scaled variance of the kinetic energy  
$\Sigma={{\sqrt  {<K^2> -<K>^2}} \over \sqrt{N}}$, are compared to the theoretical 
prediction in the microcanonical ensemble. The latter is given by the formula
$\Sigma= {T\over \sqrt{2}}  
\sqrt{  1 -{ [1-2 M({ dM\over dT })  ]}^{-1}  }$ \cite{latora99}.
Finally, in panel (d), we compute the specific heat per particle  at constant volume
in the microcanonical ensemble using the formula proposed for the first time 
by  Lebowitz, 
Percus and Verlet \cite{lpv}: 
\begin{equation} 
C_V = { 1\over 2} \left[ 1- 2 \left({ \Sigma \over T} \right)^2 \right] ^{ -1} ~~.~~~ 
\label{cv1} 
\end{equation}
%
%
\begin{figure}
\begin{center}
\epsfig{figure=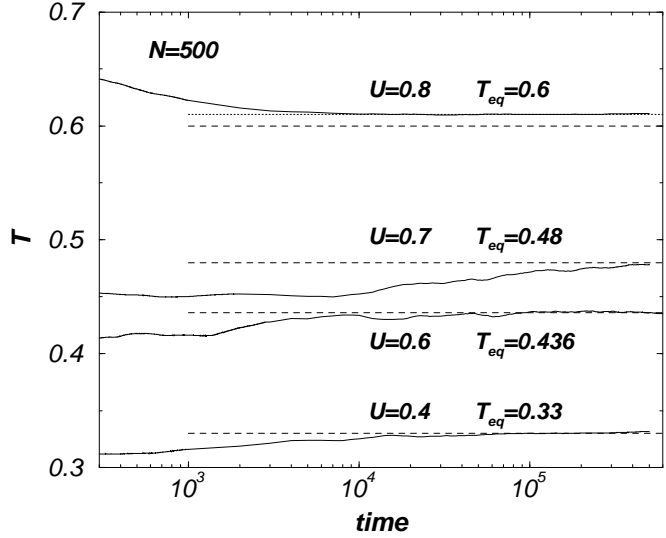,width=8truecm,angle=-90}
\end{center}
\caption{ We show, for a typical event, the 
time evolution of temperature $T$ for $N=500$ and different 
energy densities $U=0.4,0.6,0.7,0.8$. The dashed lines are the corresponding 
canonical equilibrium values. Note that for U=0.8 the dynamical temperature converges
to a slightly higher value (dotted line), which is the equilibrium one considering 
finite-size effects, see text.
}
\end{figure}

In general, the numerical microcanonical caloric curve and the  magnetization 
obtained at very long times agree quite well with 
the canonical predictions, apart from small finite-size 
effects located mainly 
around the critical point and in the supercritical region, the homogeneous phase.
In the latter,  the magnetization, due to the fact that N=500, 
is not exactly zero and assumes a finite value, i.e.   $M\sim 1/\sqrt{N}$. 
This has visible consequences on the temperature, according to formula (2) and thus also 
on the kinetic energy fluctuations and  the specific heat. 
We note  that the microcanonical specific heat at equilibrium 
is always positive and agrees with the theoretical formula (reported as a solid line).
In conclusions the canonical ensemble and the microcanonical simulations at equilibrium 
are in good agreement. 
The equivalence of different ensembles in the thermodynamic limit is also 
supported by recent analytical results in the microcanonical ensemble 
\cite{ruffod}.

\begin{figure}
\begin{center}
\epsfig{figure=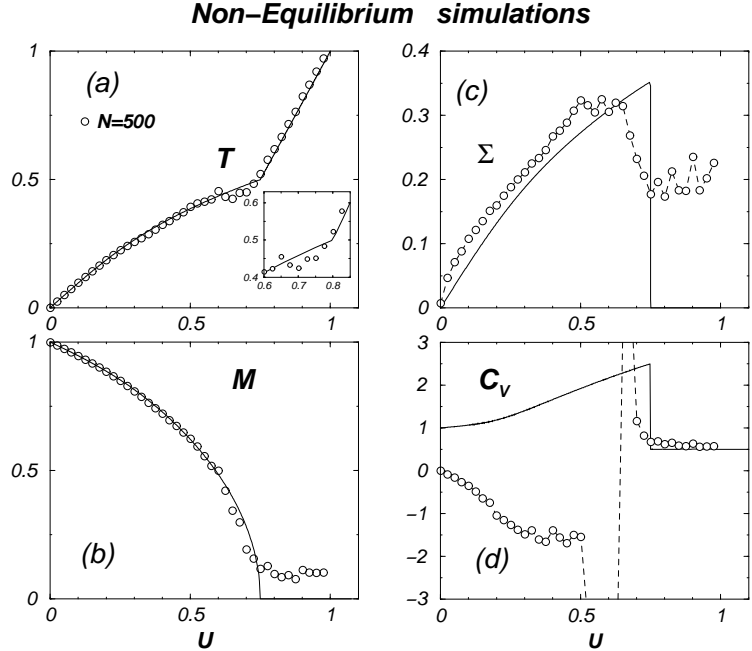,width=9truecm,angle=-90}
\end{center}
\caption{ Non-equilibrium microcanonical simulations N=500 (open circles)
in comparison with the theoretical predictions (full curves).
The integration was stopped at time $t=3000$ and the results  averaged over several
runs. Compare also with   figs. 2,4 and see   
text for further details.}
\end{figure}

\subsection{Out-of-Equilibrium results}

In the previous subsection we have presented the microcanonical numerical results 
obtained in the equilibrium regime. In the present one we 
study the process of relaxation to equilibrium and we show that HMF has 
a very rich dynamics. 

In fig.2 we report the time evolution of the temperature for 
different energies, and we study how it relaxes to the canonical value, 
represented by dashed straight lines.
Though for U=0.8,0.4 the equilibrium temperature 
is almost reached already at a time $t=3000$, close to the critical point, and in particular  
for the energy U=0.7 ,  we have to wait for much longer times 
(for a time $t> 5 \cdot  10^5$). 
In fact, before the relaxation to the canonical value, there is a 
well defined plateau, where the temperature assumes a constant 
value smaller than the canonical prediction. 
The same behavior appears in a whole energy region $0.5<U \le U_c$ and, 
as shown in ref.\cite{prl99,nostrodenton}, it corresponds to the presence of 
Quasi-Stationary States (QSS). 
The plateaus are longer the greater the system size,  
and in the $N \rightarrow \infty$  the system does not relax to the standard canonical 
equilibrium, remaining forever in the QSS \cite{nostrodenton}.
The QSS represent a different dynamical equilibrium for non-extensive systems, 
and according to an intriguing scenario recently proposed by  Tsallis \cite{tsallis},
the reason could be the inversion of the time limit 
with the size limit, an inversion which is 
implicit in our numerical simulations.

\begin{figure}
\begin{center}
\epsfig{figure=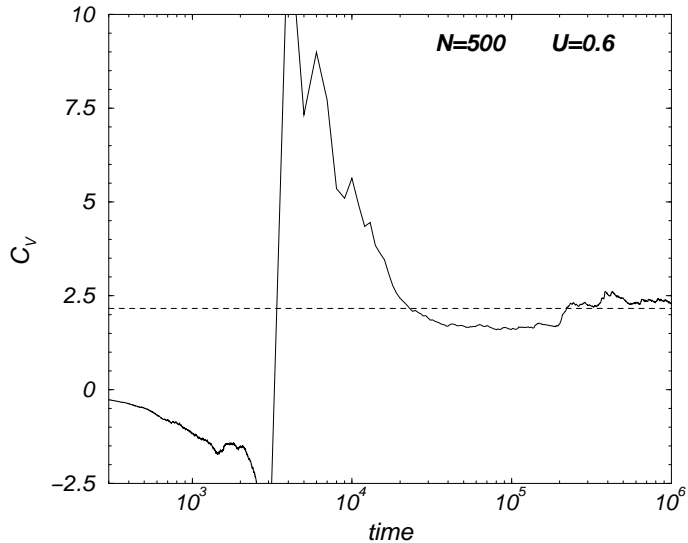,width=8truecm,angle=-90}
\end{center}
\caption{ We show, for a typical event, the 
time evolution of the specific heat per particle at constant volume $C_V$ for 
$N=500$ and $U=0.6$. $C_V$ assumes negative values in the non-equilibrium regime 
(up to time $t\sim 3000$). The dashed straight line gives the equilibrium value.}
\end{figure}

In fig.3 we report the same quantities shown in fig.1, now calculated for a very short time, i.e.
at  $t=3000$ instead of $t=2 \cdot 10^6$. For this time, 
average variables, such as $T$ and $M$, have almost reached the canonical equilibrium 
for a large range of initial energies $U$, see panel (a) and (b). 
There are, however,   relevant  differences with respect to the equilibrium  values
in the region $0.5< U \le U_c$, in correspondence to the presence of the QSS, where
the caloric curve T(U)  shows a backbending 
typical of a first-order phase transition,  see in particular the magnification reported 
in the inset of panel (a).
On the other hand,  fluctuations are affected by non-equilibrium 
effects in the whole spectrum of energy. 
In fig.3(c) the kinetic energy fluctuations are bigger than 
the equilibrium ones. As a consequence, see eq.4, the specific
 heat per particle reported   in fig.3(d)  assumes negative values 
and the behavior of $C_V$ versus U thus obtained is similar to the one found in nuclear 
multifragmentation data, see fig.4 of ref.\cite{dago}.
In particular the authors of ref.\cite{dago} claim that such a negative branch 
in the heat capacity is a direct evidence of a first-order phase transition.
Here we have a practical example that also a system with a second-order phase 
transition can show a negative specific heat 
and a backbending in the caloric curve, 
due to the non perfect equilibration. 
This effect increases with the size of the system and becomes stable in the  
limit $N \rightarrow \infty$ \cite{prox,nostrodenton}.
This dynamical effect that simulates a first-order phase transition 
can be somehow explained by the presence of superdiffusion and L\'evy walks 
in the out-of-equilibrium regime \cite{prl99}, that implies the coexistence 
of a liquid (clustered particles) and a gas (free particles) phase \cite{prox}.

It is interesting to investigate the relaxation of the specific heat  
to the equilibrium value. In fig.4 we report $C_V$, calculated by means 
of eq. \ref{cv1}, versus time for the case U=0.6.  
$C_V$ is negative in the out-of-equilibrium regime, in correspondence of QSS,
and converges very slowly to the  
equilibrium value. It is therefore necessary to wait for very long times, longer than
for the temperature thermalization, 
to have also the thermalization of  specific heat to positive values.  
A more complete study of the specific heat in HMF is in preparation\cite{prox}. 
We notice in conclusions  that the slow relaxation around the critical point
persist notwithstanding the
strong chaoticity found for the microscopic dynamics  and reported in our previous work
\cite{latora98,latora99,pro2000}. The reason for that is not completely understood,
though it has been found that the Lyapunov exponent 
is proportional to the kinetic fluctuations
 and  some heuristic conjectures have been proposed  \cite{nostrodenton}.

\section{CONCLUSIONS}

We have presented equilibrium and out-of equilibrium microcanonical simulations 
of HMF, a simple system of interacting spins.  
Though the model is not explicitly built to describe nuclear system, we think 
it can give interesting information for the multifragmentation phase transition. 
In fact, HMF can be solved in the canonical ensemble and has 
a second-order phase transition in the thermodynamic limit. 
Moreover, the advantage with respect to Ising,   lattice gas models and percolation, 
is that HMF  
allows a dynamical microcanonical approach to the study of phase transitions 
in a finite system.  
HMF has a very rich dynamical behavior in a transient out-of-equilibrium 
regime whose timescale depends on the energy and on the size of the system.  
When the microcanonical simulations are started in an out-of-equilibrium 
initial state, for example in a ``water bag'', we find the appearance of 
quasi-stationary-states.  
In correspondence we have a caloric curve with a well defined backbending and 
negative specific heat.  
These indications of a first-order phase transition, at variance with the 
second-order phase transition predicted in the thermodynamic limit, 
are effects of the non-equilibrium. 
Certainly the non-equilibrium features of the model are strictly linked to 
the long-range nature of the interaction, however the  general validity of 
these results   is not completely clear and further work in this direction is needed.

\bigskip
We thank X. Campi,
M. Pettini and C. Tsallis for  useful discussions.
Part of this work has been done in collaboration with S. Ruffo.


\end{document}